\documentclass[pra,aps,twocolumn,superscriptaddress,floatfix,showpacs,longbibliography]{revtex4-2}
\usepackage{graphicx}
\usepackage{dcolumn}
\usepackage{bm}
\usepackage{upgreek}
\usepackage{natbib}
\usepackage[normalem]{ulem}
\usepackage{amsmath,amssymb}
\usepackage{dsfont}
\usepackage[english]{babel}
\usepackage{braket}
\usepackage{diagbox}
\usepackage{color}

\begin{document}
\title{Moir\'e and frustration physics of dipolar supersolids under periodic confinement} 
\author{Ze-Hong Guo}
\thanks{These authors contributed equally to this work.}
\affiliation{Key Laboratory of Atomic and Subatomic Structure and Quantum Control (Ministry of Education), Guangdong Basic Research Center of Excellence for Structure and Fundamental Interactions of Matter, School of Physics, South China Normal University, Guangzhou 510006, China}
\author{Kai Gan}
\thanks{These authors contributed equally to this work.}
\affiliation{Key Laboratory of Atomic and Subatomic Structure and Quantum Control (Ministry of Education), Guangdong Basic Research Center of Excellence for Structure and Fundamental Interactions of Matter, School of Physics, South China Normal University, Guangzhou 510006, China}
\author{Qizhong Zhu}
\email{qzzhu@m.scnu.edu.cn}
\affiliation{Key Laboratory of Atomic and Subatomic Structure and Quantum Control (Ministry of Education), Guangdong Basic Research Center of Excellence for Structure and Fundamental Interactions of Matter, School of Physics, South China Normal University, Guangzhou 510006, China}
\affiliation{Guangdong Provincial Key Laboratory of Quantum Engineering and Quantum Materials, Guangdong-Hong Kong Joint Laboratory of Quantum Matter, Frontier Research Institute for Physics, South China Normal University, Guangzhou 510006, China}

\date{\today}

\begin{abstract}
We study the ground-state phases of a two-dimensional dipolar supersolid subjected to external periodic confinement by numerically solving the extended Gross--Pitaevskii equation. Focusing on a regime in which the unconfined system forms an intrinsic triangular droplet crystal, we consider triangular, honeycomb, and square optical lattices and classify them into isostructural and heterostructural settings relative to the spontaneous supersolid order. We map out the stationary states as functions of the lattice depth $V_0$ and the commensurability ratio between the intrinsic droplet spacing and the external lattice period. For triangular and honeycomb confinements, the competition between the soft self-organized supersolid lattice and the rigid external potential can generate long-wavelength moir\'e superstructures in the weak- to intermediate-lattice regime, together with a sequence of reconstructed states including ring-like clusters and stripe-segment configurations. By contrast, the square lattice introduces strong symmetry mismatch between the intrinsic $C_6$ order and the imposed $C_4$ geometry, leading to frustration-induced anisotropic states and symmetry-reduced cluster arrangements. Our results establish dipolar supersolids under periodic confinement as an unconventional route to exploring moir\'e physics, where moir\'e superstructures arise from the competition between a self-organized soft lattice and an externally imposed rigid one.
\end{abstract}

\maketitle

\section{Introduction}
Supersolid phases, which combine global phase coherence with spontaneously broken translational symmetry, have long attracted interest in quantum many-body physics and have now been realized in several ultracold-gas platforms, including cavity-mediated, spin-orbit-coupled, and dipolar systems \cite{boninsegni2012colloquium,chomaz2016quantum,baumann2010dicke,leonard2017supersolid,leonard2017monitoring,guo2021optical,li2017stripe,tanzi2019observation,bottcher2019transient,chomaz2019long}. Among these, dipolar Bose--Einstein condensates provide a controllable setting for studying self-organized quantum matter, because long-range and anisotropic dipole-dipole interactions compete with kinetic energy, short-range interactions, and quantum fluctuations on comparable footing \cite{giovanazzi2002tuning,lahaye2009physics,defenu2023long,lu2011strongly,aikawa2012erbium,bottcher2020new,chomaz2025quantum,ferrier2016observation,chomaz2023dipolar,lee1957eigenvalues}. In the regime where the tendency toward roton softening is arrested by beyond-mean-field stabilization, the condensate can spontaneously develop spatial density modulation while retaining global phase coherence, thereby realizing a dipolar supersolid \cite{santos2003roton,kadau2016rosensweig,schmitt2016self,chomaz2018roton,lu2015stable,petrov2015quantum,lima2011quantum,lima2012beyond,wachtler2016ground,bisset2016ground,zhang2019supersolidity,bottcher2019transient,chomaz2019long,tanzi2019observation}. Recent work has further established that dipolar quantum gases support a rich family of such states, including stripe, triangular, and honeycomb supersolids in two dimensions \cite{wenzel2017striped,zhang2021phases,hertkorn2021pattern,norcia2021two,bland2022two,ripley2023two}. A defining feature of these phases is that their periodicity is not externally imposed, but selected self-consistently by interactions. The resulting crystal is therefore soft, compressible, and highly tunable.

This spontaneous crystalline order immediately suggests a broader question: how does a dipolar supersolid respond when it is subjected to an additional periodic environment? Optical lattices provide a controllable way to address this problem \cite{matsuda1970off,baier2016extended,su2023dipolar}. In a weakly interacting condensate, an external lattice mainly acts as a rigid background that reshapes the single-particle dispersion and pins the density. Related studies in spin-orbit-coupled Bose gases have shown that periodic potentials can pin one-dimensional (1D) stripe supersolids and qualitatively reshape their low-energy excitations \cite{li2021pseudo,guo2024zn}. For dipolar supersolids, however, the interplay is more intricate because the spontaneous order is a two-dimensional (2D) droplet crystal, so the interplay with the geometry of the external optical lattice becomes essential. The system already carries an intrinsic lattice scale and symmetry through its droplet crystal, so the applied optical potential introduces a second, externally fixed periodic structure. The problem is therefore not simply that of loading a Bose gas into a lattice, but rather that of confronting a soft interaction-generated crystal with a rigid imposed one.

Viewed from this perspective, periodic confinement of a dipolar supersolid provides a cold-atom setting for studying moir\'e-like effects and geometric frustration. In conventional moir\'e materials, emergent long-wavelength structures arise from the competition between two rigid periodic lattices with slightly different periods or orientations \cite{geim2013van,hunt2013massive,dean2013hofstadter,cao2018correlated,cao2018unconventional,lu2019superconductors,cao2021nematicity,zeng2023thermodynamic,xu2023observation,balents2020superconductivity,kennes2021moire}. Closely related ideas have been extended from van der Waals heterostructures to synthetic lattices and other engineered quantum platforms, where geometric mismatch can strongly reshape single-particle spectra and correlated many-body states \cite{gonzalez2019cold,luo2021spin,meng2023atomic,zeng2025interaction,tian2025dynamics,ding2025interaction,wang2024threedimensional,wan2026fractal}. Aperiodic and quasicrystalline optical potentials provide a related cold-atom setting in which competing length scales already generate localization, fractality, and unconventional Bose-glass behavior \cite{viebahn2019matter,yao2019critical,yao2020lieb,zhu2023thermodynamic,ciardi2023quasicrystalline,yu2024observing}. In the present setting, one of the two competing structures is instead generated dynamically by interactions and can continuously deform in response to the external field. This added softness makes the competition intrinsically nonlinear and self-consistent: relative symmetry, commensurability, and lattice depth can drive pinning, droplet splitting, connectivity changes, or rotational-symmetry breaking, and may also produce long-wavelength moir\'e-like density modulations.

Despite rapid progress in both dipolar supersolids and lattice engineering, the interplay between rigid periodic confinement and 2D dipolar droplet supersolids has received limited attention. Related studies of confined dipolar supersolids and of dipolar gases in aperiodic optical environments have already revealed multiple self-organized states, highlighting the role of externally designed confinement in long-range interacting bosonic systems \cite{zhang2021phases,zampronio2024exploring,viebahn2019matter,yao2020lieb,zhu2023thermodynamic,yu2024observing}. Several basic questions therefore follow. When the external lattice is symmetry-compatible with the intrinsic droplet crystal, does the supersolid evolve smoothly into a pinned lattice state, or does it pass through intermediate reconstructions? When the two structures are symmetry-incompatible, how is geometric frustration relieved? Under what conditions do emergent moir\'e superstructures appear, and when do anisotropic stripe-like states become energetically favorable? Answering these questions is essential for understanding how tailored confinement can be used to control the symmetry, connectivity, and characteristic length scales of supersolid order.

In this work, we investigate these issues for a 2D dipolar Bose--Einstein condensate in the supersolid regime subjected to triangular, honeycomb, and square optical lattices. We focus on a parameter window in which the unperturbed system forms an intrinsic triangular droplet crystal, so that the applied lattices naturally separate into an isostructural case (triangular confinement) and heterostructural cases (honeycomb and square confinement). By solving the extended Gross--Pitaevskii equation (eGPE) for a broad set of initial states, we map out the stationary density configurations as functions of the lattice depth $V_0$ and the commensurability ratio between the intrinsic droplet spacing and the external lattice period. We show that the competition between the soft self-organized supersolid lattice and the rigid optical confinement generates a sequence of reconstructions. In the symmetry-matched triangular case, the system evolves mainly within a sixfold-symmetric manifold and can support moir\'e-modulated states as well as intermediate cluster reconstructions. In the honeycomb case, the additional geometric mismatch and local repulsion drive stronger restructuring, including stripe-segment states and loop- or ring-like droplet arrangements, while commensurate settings can still support moir\'e superlattices. In the square case, the incompatibility between the intrinsic sixfold ($C_6$) order and the imposed fourfold ($C_4$) symmetry produces frustration-induced anisotropic states and symmetry-reduced cluster patterns. These results establish periodic confinement as a route for engineering frustrated supersolid matter and identify dipolar supersolids under periodic confinement as an unconventional route to moir\'e physics, where moir\'e superstructures arise from the competition between a self-organized soft lattice and an externally imposed rigid one.

\section{Model and Method}
\subsection{System and three-dimensional eGPE}
\begin{figure}[tbp] \centering
	\includegraphics[width=0.99\linewidth]{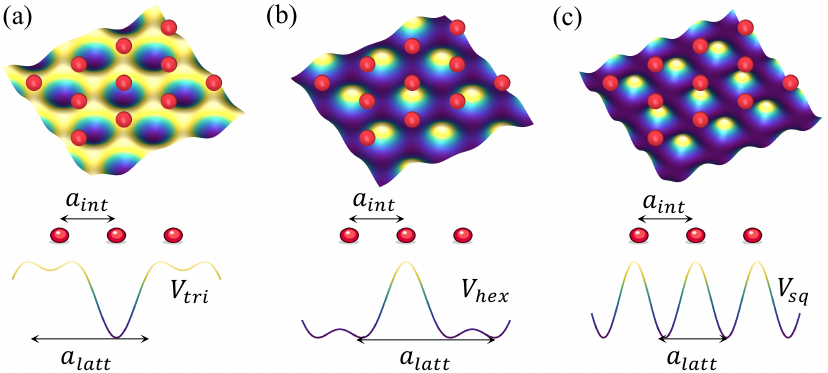}
	\caption{(a)-(c) Schematic illustrations of the system under study. A 2D dipolar supersolid (orange spheres forming an intrinsic triangular array) is subjected to an external periodic optical potential $V_{\mathrm{ext}}$ with (a) triangular, (b) honeycomb, and (c) square geometry. The competition between the intrinsic droplet spacing $a_{\mathrm{int}}$ and the imposed lattice period $a_{\mathrm{latt}}$ generates commensurability effects and geometric frustration.}
	\label{fig1}
\end{figure}

We consider a Bose-Einstein condensate of $N$ $^{164}$Dy atoms of mass $m$, with all dipole moments polarized along the $z$ axis \cite{lu2011strongly}. The system is tightly confined by a harmonic trap in the axial direction, $V_z(z)=m\omega_z^2 z^2/2$, and subjected in the $(x,y)$ plane to a periodic optical potential $V_{\mathrm{ext}}(\mathbf r)$, where $\mathbf r=(x,y)$ and $\mathbf R=(\mathbf r,z)$ denote the full three-dimensional (3D) coordinate [Fig.~\ref{fig1}]. 
Throughout this work, the condensate wave function is normalized as $\int d\mathbf R\,|\Psi(\mathbf R,t)|^2=N$,  where $N$ is the total particle number.

At the level of the extended Gross-Pitaevskii description with local-density quantum-fluctuation corrections 
\cite{petrov2015quantum,lima2011quantum,wachtler2016ground,bisset2016ground,bottcher2020new}, the 3D eGPE reads
\begin{equation}
{\begin{aligned}
i\hbar\frac{\partial\Psi}{\partial t}={}&\bigg[-\frac{\hbar^2\nabla^2}{2m}+\frac{1}{2}m\omega_z^2z^2+V_{\mathrm{ext}}(\mathbf r)+g_s|\Psi|^2\\
&+\gamma_{\mathrm{QF}}|\Psi|^3+\int d\mathbf{R}^\prime\,V_{\mathrm{dd}}(\mathbf{R}-\mathbf{R}^\prime)|\Psi(\mathbf{R}^\prime)|^2\bigg]\Psi,
\end{aligned}}
\end{equation}
where $\nabla^2$ is the 3D Laplacian. 
Here 
\begin{equation}
\gamma_{\mathrm{QF}} = \frac{128\sqrt{\pi}\hbar^2 a_s^{5/2}}{3m} \left[ 1 + \frac{3}{2}\left(\frac{a_{\mathrm{dd}}}{a_s}\right)^2 \right]
\end{equation}
is the LHY fluctuation coefficient \cite{petrov2015quantum,lima2011quantum}, and $V_{\mathrm{dd}}(\mathbf{R})=\frac{\mu_0\mu_m^2}{4\pi}\frac{1-3\cos^2\theta}{R^3}$ is the dipole-dipole interaction kernel, while $g_s={4\pi\hbar^2 a_s}/m$ denotes the contact-interaction strength. $a_s$ is the $s$-wave scattering length and $a_{\mathrm{dd}}$ is the dipolar strength.
The corresponding mean-field energy functional is 
\begin{equation}
{\begin{aligned}
E=&\int d\mathbf{R} \, \Psi^*(\mathbf R) \left[ -\frac{\hbar^2 \nabla^2}{2m} + \frac{1}{2}m\omega_z^2 z^2 + V_{\mathrm{ext}}(\mathbf r) \right] \Psi(\mathbf R)\\
&+\frac{1}{2}\int d\mathbf{R}\,d\mathbf{R}^\prime\,V_{\mathrm{dd}}(\mathbf{R}-\mathbf{R}^\prime)|\Psi(\mathbf R)|^2|\Psi(\mathbf R^\prime)|^2\\
&+\frac{g_s}{2}\int d\mathbf{R}\,|\Psi(\mathbf R)|^4+\frac{2}{5}\gamma_{\mathrm{QF}}\int d\mathbf{R}\,|\Psi(\mathbf R)|^5.
\end{aligned}}
\label{3Dfunctional}
\end{equation}

\subsection{Quasi-two-dimensional reduction}
Because the confinement along $z$ is strong, we adopt a factorized 3D ansatz, $\Psi(\mathbf r,z,t)=\sqrt{n}\psi(\mathbf r,t)\chi(z)$, where $n$ denotes the areal density of particles in the $(x,y)$ plane, $\psi(\mathbf r,t)$ is the in-plane wave function, and $\chi(z)$ describes the axial mode.
The reduced wave functions satisfy $\int d\mathbf r\,|\psi(\mathbf r,t)|^2=A$ and $\int dz\,|\chi(z)|^2=1$, where $A$ stands for the area of the system. For the axial mode, we use the normalized Gaussian form $\chi(z)=\exp\!\left(-z^2/(2\sigma_z^2)\right)/(\pi\sigma_z^2)^{1/4}$,
where $\sigma_z$ is treated variationally. 
Substituting this ansatz into Eq.~(\ref{3Dfunctional}) and integrating out the axial degree of freedom yields the effective 2D functional
\begin{equation}
\begin{aligned}
E[\psi]=&n\int d\mathbf{r}\,\psi^*(\mathbf{r})\left[-\frac{\hbar^2\nabla^2}{2m}+V_{\mathrm{ext}}(\mathbf r)+\mathcal E_z\right]\psi(\mathbf{r})\\
&+\frac{g_{\mathrm{2D}}n^2}{2}\int d\mathbf{r}\,|\psi(\mathbf{r})|^4+\frac{n^2}{2}\int d\mathbf{r}\,\Phi_{\mathrm{dd}}(\mathbf{r})|\psi(\mathbf{r})|^2\\
&+\frac{2}{5}g_{\mathrm{QF}}n^{5/2}\int d\mathbf{r}\,|\psi(\mathbf{r})|^5,
    \end{aligned}
    \label{eq:2Dfunctional_clean}
\end{equation}
where $\mathcal E_z=\hbar^2/(4m\sigma_z^2)+m\omega_z^2\sigma_z^2/4$ is the axial single-particle contribution, $g_{\mathrm{2D}}=g_s/(\sqrt{2\pi}\sigma_z)$ is the effective contact interaction strength, $g_{\mathrm{QF}}=\gamma_{\mathrm{QF}}\sqrt{2/5}(1/\pi\sigma_z^2)^{3/4}$ is the quasi-2D LHY coefficient. The effective dipolar interaction potential is obtained via convolution, $\Phi_{\mathrm{dd}}(\mathbf{r})=\mathcal{F}^{-1}\left[\widetilde{U}(k_{\rho})\,\mathcal{F}\left(|\psi(\mathbf{r})|^{2}\right)\right]$, where $\mathcal{F}$ denotes the 2D Fourier transform. Here, $\widetilde{U}(k_{\rho})=\dfrac{\mu_{0}\mu_{m}^{2}}{3\sqrt{2\pi}\sigma_{z}}\left[2-3\sqrt{\pi}Q \exp{(Q^{2})}\operatorname{erfc}(Q)\right]$ is the dipolar interaction kernel in momentum space, with $Q=k_{\rho}\sigma_{z}/\sqrt{2}$ and $k_{\rho}=\sqrt{k_{x}^{2}+k_{y}^{2}}$ \cite{ticknor2011anisotropic,fischer2006stability,ripley2023two}. In the following simulations, we use $l=12\pi a_{\mathrm{dd}}$ as the unit of length and $E_0=\hbar^2/(m l^2)$ as the unit of energy.

\subsection{External periodic potentials}
To probe the response of the dipolar supersolid to externally imposed geometry, we consider the model optical potential
\begin{equation}
V_{\mathrm{ext}}(\mathbf r)=sV_0\sum_{j=1}^{M}\cos(\mathbf k_j\cdot\mathbf r),
\label{eq:Vext_clean}
\end{equation}
where $V_0>0$ is the lattice depth, $s=\pm1$ selects the sign of the modulation, and $\{\mathbf k_j\}_{j=1}^{M}$ is a set of in-plane reciprocal vectors of equal magnitude $|\mathbf k_j|=k_L$ that defines the lattice geometry. We consider three representative choices of this set.

(i) Triangular lattice. For $M=3$, with reciprocal vectors separated by $120^\circ$ and generated by successive counterclockwise rotations of $\mathbf k_1=k_L(0,1)$, choosing $s=-1$ produces a triangular potential with lattice constant $a_{\mathrm{latt}}=4\pi/(\sqrt{3}\,k_L)$ \cite{becker2010ultracold,struck2011quantum}.

(ii) Honeycomb lattice. Using the same three-vector set but choosing $s=+1$ in Eq.~(\ref{eq:Vext_clean}) yields a honeycomb potential \cite{soltan2011multi,tarruell2012creating,jotzu2014experimental}.

(iii) Square lattice. For $M=2$, with $\mathbf k_1=k_L\hat{\mathbf e}_x$ and $\mathbf k_2=k_L\hat{\mathbf e}_y$, we obtain a square potential with lattice constant $a_{\mathrm{latt}}=2\pi/k_L$ \cite{wirth2011evidence}.

Based on the structural and symmetry relations between the external potential and the unperturbed triangular supersolid, we classify the systems into isostructural and heterostructural cases. The triangular potential is isostructural: it is symmetry matched to the intrinsic droplet crystal, so the competition is governed primarily by the relative length scales and energy scales. By contrast, the honeycomb and square potentials are heterostructural and introduce geometric frustration. The honeycomb lattice preserves the global $C_6$ symmetry but places potential maxima at the positions preferred by the intrinsic droplets, thereby generating a repulsive local frustration. The square lattice imposes an incompatible $C_4$ symmetry on the triangular supersolid and therefore produces a stronger form of geometric frustration.

\subsection{Parameters and numerical procedure}
We adopt parameters appropriate to $^{164}$Dy condensates \cite{lu2011strongly,norcia2021two,bland2022two}. The axial confinement is fixed at $\omega_z=2\pi\times72.4\,\mathrm{Hz}$, and the dipolar length is $a_{\mathrm{dd}}=130.8\,a_0$, where $a_0$ is the Bohr radius. In dimensionless form, we choose $(na_{\mathrm{dd}}^2, a_s/a_{\mathrm{dd}})=(0.0978,0.755)$ \cite{ripley2023two}. Under these conditions, the unperturbed system at $V_0=0$ forms a robust triangular supersolid. By comparing the energies of stationary states with different lattice spacings, we determine the optimized intrinsic droplet spacing to be $a_{\mathrm{int}}\approx17.86\,l$, which then serves as the reference scale for all commensurability analyses.

The modulated ground states are obtained by minimizing Eq.~(\ref{eq:2Dfunctional_clean}) through imaginary-time evolution with periodic boundary conditions. Because the competition between the intrinsic dipolar interactions and the external lattice generates many metastable solutions, the numerical evolution can become trapped in different local minima. To identify the lowest-energy converged state as reliably as possible, we initialize the evolution from several distinct trial states: a uniform state, a 1D stripe state $\psi(\mathbf r)\propto\sum_i \exp\!\left[-(x-x_i)^2/(2\sigma)^2\right]$, and 2D  droplet-crystal states (triangular or honeycomb) $\psi(\mathbf r)\propto\sum_i \exp\!\left[-\big((x-x_i)^2+(y-y_i)^2\big)/(2\sigma^2)\right]$, where $\sigma$ sets the initial in-plane droplet width. For each parameter set, the state with the lowest converged energy is identified as the ground state, while the axial width $\sigma_z$ is optimized simultaneously during the minimization.

In this work, to satisfy periodic boundary conditions for the intrinsic supersolid lattice, the computational domain is taken as $[-L_x,L_x]\times[-L_y,L_y]$, with $L_x=3a_{\mathrm{int}}$ and $L_y=3\sqrt{3}\,a_{\mathrm{int}}$, and discretized on a $256\times256$ grid. In the numerical calculations, the imaginary-time evolution is considered converged when the energy difference between two successive checks, performed every 4000 steps, falls below $10^{-7} E_0$.

The problem is controlled by three parameters: the lattice depth $V_0$, the commensurability ratio $R=a_{\mathrm{int}}/a_{\mathrm{latt}}$, and the lattice geometry. Varying $V_0$ tunes the energetic competition between the rigid external modulation and the intrinsic interaction-driven ordering tendency, thereby driving the system from the weak-lattice regime through intermediate reconstruction regimes toward the deep-lattice limit. 
Deviating $R$ from unity introduces a controlled mismatch between the intrinsic and imposed length scales. This translational incompatibility is the key ingredient for the emergence of long-wavelength moir\'e-like structures, generated by the beating between $a_{\mathrm{int}}$ and $a_{\mathrm{latt}}$. Changing the lattice geometry allows us to compare symmetry-compatible and symmetry-incompatible constraints, and hence to distinguish the effects of commensurability from those of geometric frustration acting on the same underlying supersolid.

Note that throughout this work, the applied lattice depths are kept within the superfluid regime, so the relevant physics is the reorganization of superfluid density rather than the onset of Mott localization. Furthermore, the presence of an external potential explicitly breaks translational symmetry. Therefore, the system's ground state is no longer a supersolid in the strict sense, which is characterized by the spontaneous breaking of both U(1) gauge symmetry and translational symmetry \cite{li2021pseudo,guo2024zn}. For this reason, we refer to the ground state under the confinement of an external periodic potential as a superfluid rather than a supersolid.

\section{Results}

The response of the dipolar supersolid to an optical lattice is controlled by two parameters: the symmetry relation between the intrinsic triangular droplet crystal and the imposed potential, and the commensurability ratio $R=a_{\rm int}/a_{\rm latt}$. For the triangular lattice, which is isostructural with the unperturbed supersolid, the evolution is comparatively smooth and remains largely confined to the sixfold-symmetric sector, except when a short lattice period transiently favors stripe-like states. By contrast, the honeycomb and square lattices are heterostructural. In those cases, symmetry mismatch and geometric frustration produce a denser spectrum of metastable branches and a broader sequence of intermediate density reconstructions. Throughout this section, the relaxed-energy branches obtained from different initial conditions are used to identify the ground state, and the associated density profiles are then interpreted in terms of the competition between dipolar self-organization and external confinement.

\subsection{Isostructural case}

\begin{figure}[tbp]
    \centering
    \includegraphics[width=0.99\linewidth]{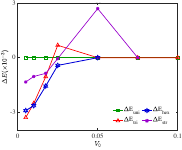} 
    \caption{Energies of converged stationary branches obtained from different initial states for the system subjected to a triangular potential, with a fixed periodicity ratio $R=1/2$. 
    Each curve corresponds to a different initial configuration including uniform (uni), triangular (tri), hexagonal (hex), and stripe (str).
    Energies are measured relative to the relaxed uniform state, so $\Delta_{\rm uni}=0$.}
    \label{fig2}
\end{figure}

We first consider the triangular optical lattice, which preserves the $C_6$ symmetry of the intrinsic droplet crystal. Symmetry matching suppresses strong frustration in this case. For $R<1$, the ground state evolves mainly through continuous reconstruction within a sixfold-symmetric manifold; for $R>1$, the shorter lattice period can temporarily favor anisotropic states, but the deep-lattice configuration is still selected by the triangular potential.

This behavior is already evident from the representative energy landscape at $R=1/2$ in Fig.~\ref{fig2}. At weak lattice depth, the triangular and hexagonal initial states relax to the lowest energies, indicating that the external field primarily pins and modulates the intrinsic supersolid rather than destabilizing it. The stripe branch remains high in energy over the entire range shown, which confirms that 1D ordering is strongly disfavored when the external confinement respects the intrinsic symmetry. At larger $V_0$, the energy branches move closer together, signaling the growing dominance of the imposed lattice.

\begin{figure}[tbp]
    \centering
    \includegraphics[width=0.99\linewidth]{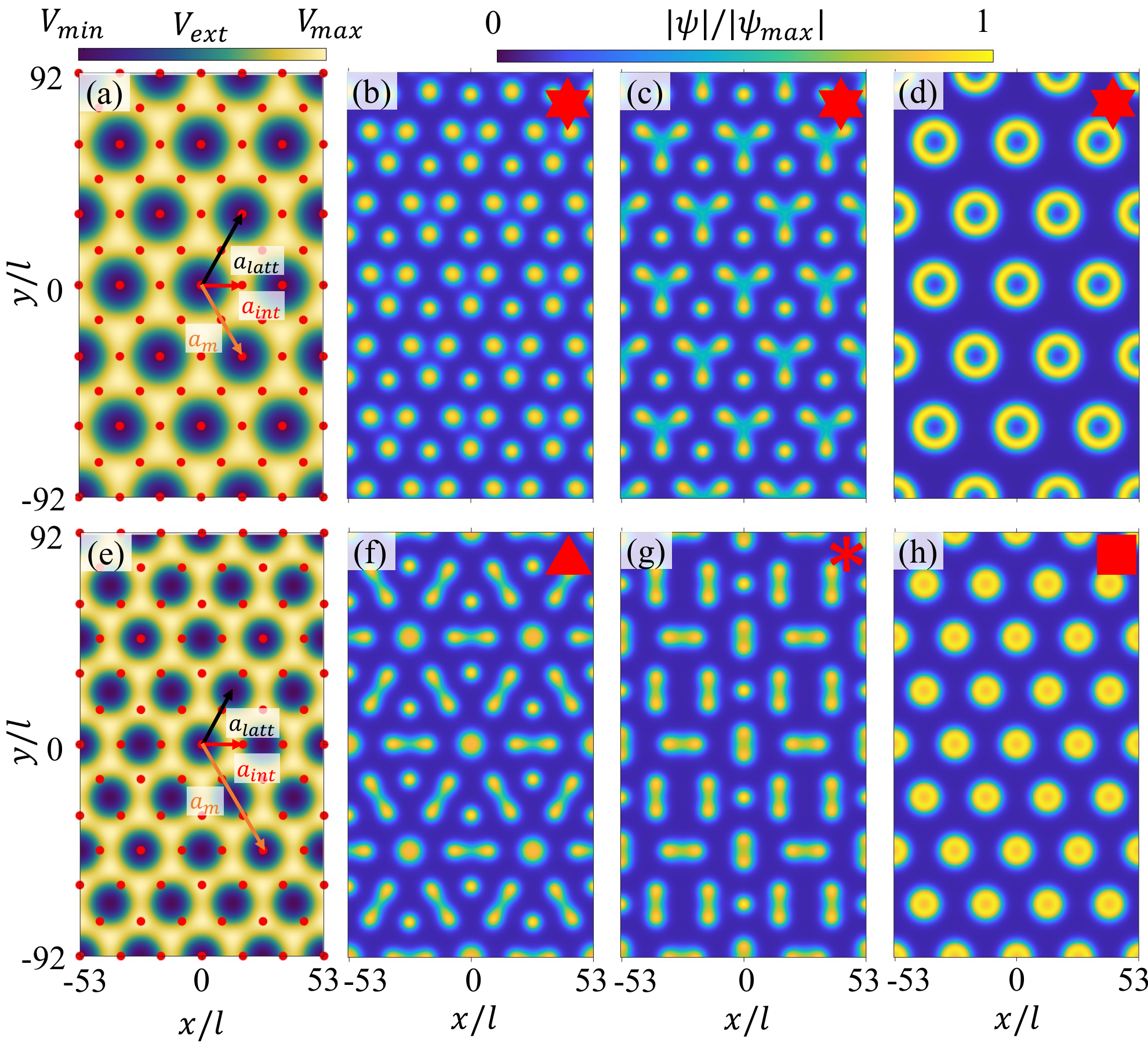}
    \caption{Ground-state density evolution of the dipolar Bose gas in a triangular optical lattice for $R < 1$.
    (a), (e) Superposition of the intrinsic triangular density (red dots) and the external potential (color map) for $R = 1/2$ and $R = 2/3$, respectively.
    The arrows mark the three competing length scales: $a_{\text{int}}$ (red), $a_{\text{pot}}$ (black), and the emergent moir\'e period $a_m$ (orange).
    (b)--(d) Density profiles for $R = 1/2$ at increasing lattice depths $V_0 = 0.0175$, $0.025$, and $0.05$, respectively.
    (f)--(h) Density profiles for $R = 2/3$ at increasing lattice depths $V_0 = 0.01$, $0.015$, and $0.02$, respectively.
    In each density panel, a geometric marker indicates the initial state from which the displayed ground-state density is obtained, using the same symbol convention as in Fig.~\ref{fig2}.
    The star in panels (b)--(d) corresponds to the hexagonal initial state, the triangle in (f), the asterisk in (g), and the square in (h) indicate the respective initial states that yield the lowest energy at each $V_0$.}
    \label{fig3}
\end{figure}

We now turn to the $R<1$ density evolution shown in Fig.~\ref{fig3}. Figures~\ref{fig3}(a) and \ref{fig3}(e) illustrate the geometric relation between the intrinsic triangular order and the imposed triangular lattice for $R=1/2$ and $R=2/3$, respectively. Their superposition, characterized by the intrinsic spacing $a_{\mathrm{int}}$ (red arrow) and the external lattice period $a_{\mathrm{latt}}$ (black arrow), generates a moir\'e superlattice with a longer period $a_m$ (orange arrow). In this regime, the moir\'e period remains commensurate with the external lattice, taking the values $a_m=a_{\mathrm{latt}}$ for $R=1/2$ and $a_m=3a_{\mathrm{latt}}$ for $R=2/3$.

For $R=1/2$ [Figs.~\ref{fig3}(b)--\ref{fig3}(d)], the same low-energy branch remains selected across the entire sequence. This is indicated directly by the identical star markers in all three panels, showing that the ground state continues to originate from the same hexagonal initial-state branch as $V_0$ increases. At weak lattice depth, the state is a moir\'e-modulated superfluid in which the droplet amplitudes are alternately enhanced and reduced while the triangular arrangement is preserved. At intermediate depth, neighboring droplets merge along the potential valleys and form trefoil-like clusters, indicating a connectivity reconstruction within the moir\'e unit cell. At larger $V_0$, these clusters separate into ring-like droplets centered at the triangular-lattice minima. The overall evolution is therefore comparatively smooth and remains within the sixfold-symmetric sector.

For $R=2/3$ [Figs.~\ref{fig3}(f)--\ref{fig3}(h)], the larger moir\'e cell supports a more competitive energy landscape. In contrast to the uniform star markers for $R=1/2$, the markers now change from triangle to asterisk to square across the three panels, signaling that the ground state {changes between different low-energy branches} as $V_0$ increases. The weak-lattice state in panel (f), marked by the triangle, is obtained from the triangular branch and remains a modulated triangular superfluid. In the intermediate regime, however, the optimal branch changes, and the density is channeled into stripe-like segments with reduced rotational symmetry. This anisotropic texture reflects a compromise between lattice-induced channeling and dipolar repulsion. At still larger $V_0$, the stripes break into discrete ring-like droplets locked to the external minima; the square marker in panel (h) shows that the ground state has switched again, now to the branch connected to the uniform initial state. Compared with $R=1/2$, the route to the lattice-dominated regime therefore proceeds through a distinct symmetry-broken intermediate state accompanied by branch switching.

\begin{figure}[tbp]
    \centering
    \includegraphics[width=0.99\linewidth]{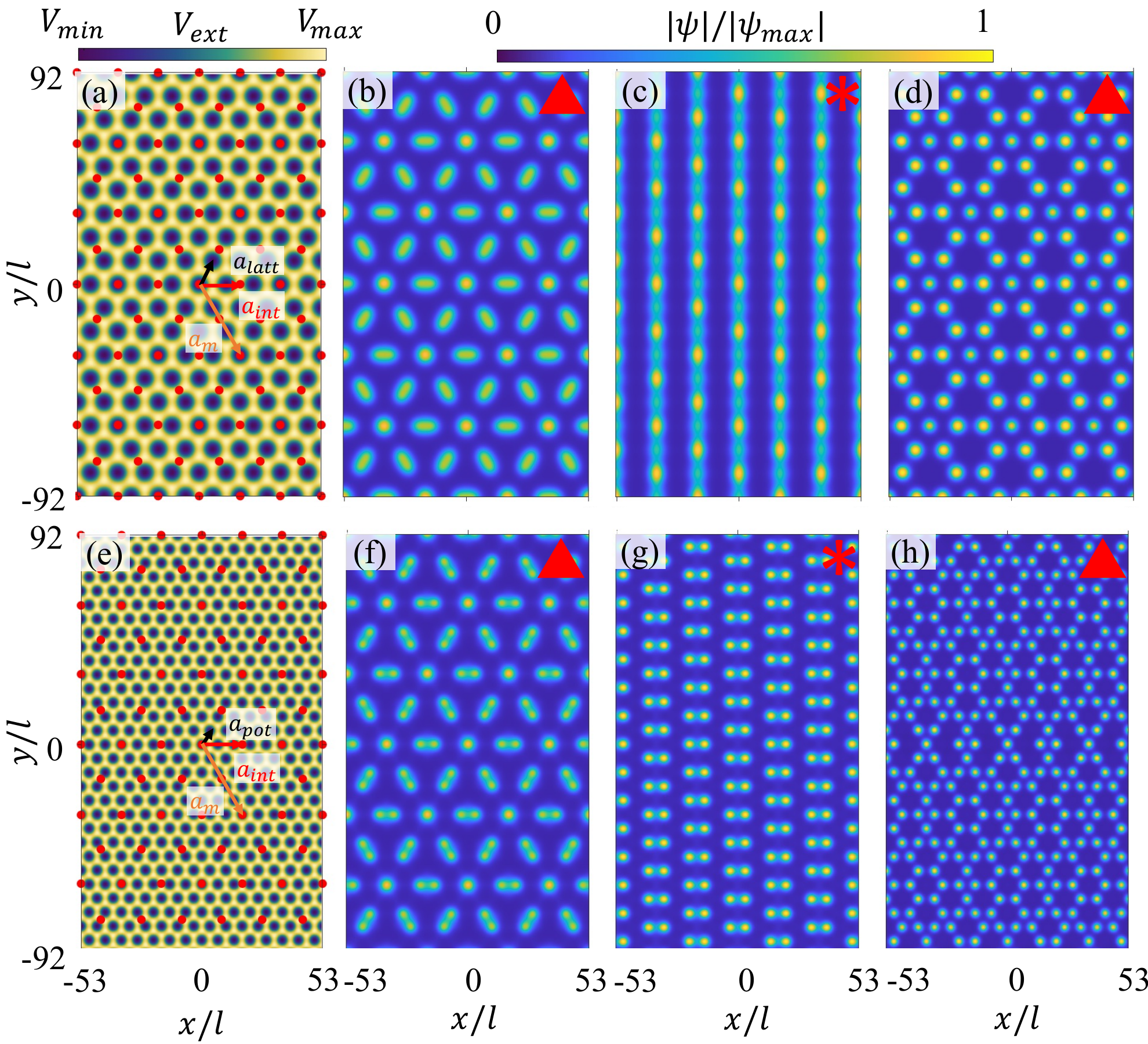}
    \caption{Ground-state density evolution of the dipolar Bose gas in a triangular optical lattice for $R > 1$.
    (a), (e) Superposition of the intrinsic triangular density (red dots) and the external potential (color map) for $R = 3/2$ and $R = 5/2$, respectively.
    The arrows indicate the three relevant length scales: $a_{\rm int}$ (red), $a_{\rm pot}$ (dark), and the moir\'e period $a_m$ (orange).
    (b)--(d) Density profiles for $R = 3/2$ at increasing lattice depths $V_0 = 0.005$, $0.0075$, and $0.025$, respectively.
    (f)--(h) Density profiles for $R = 5/2$ at increasing lattice depths $V_0 = 0.025$, $0.05$, and $0.125$, respectively.
    Intermediate lattice depths induce symmetry-broken stripe-like states: a clearly developed stripe-like state for $R = 3/2$ and a paired-droplet or multi-scale stripe texture for $R = 5/2$.
    In each density panel, the geometric marker identifies the initial state yielding the ground-state density, using the same symbol convention as in Fig.~\ref{fig2}.}
    \label{fig4}
\end{figure}

For $R>1$, the optical lattice period is shorter than the intrinsic droplet spacing, so the external field acts as a sub-droplet corrugation. Figure~\ref{fig4} shows that this regime can transiently favor anisotropic states even in the isostructural geometry. For $R=3/2$ [Figs.~\ref{fig4}(b)--\ref{fig4}(d)], the weak-lattice state remains a slightly modulated triangular superfluid. At intermediate depth, however, the stripe branch becomes the ground state and the density reorganizes into quasi-one-dimensional channels along $\hat y$, revealing spontaneous rotational-symmetry breaking. At larger $V_0$, the triangular branch recovers the lowest energy and the stripes fragment into a denser droplet array controlled mainly by the external lattice.

A similar but more elaborate sequence appears for $R=5/2$ [Figs.~\ref{fig4}(f)--\ref{fig4}(h)]. The weak-lattice state is again moir\'e modulated, whereas the intermediate regime favors a multi-scale stripe texture that combines 1D channeling with residual internal droplet structure. At larger $V_0$, the system returns to a triangular-branch configuration and evolves toward a fragmented lattice-dominated state. Thus, even in the symmetry-matched case, sufficiently short lattice periods can drive a clearly developed anisotropic intermediate regime before complete pinning is achieved.

\subsection{Heterostructural case I: honeycomb potential}

The honeycomb lattice produces a different reconstruction sequence because its maxima coincide with the positions preferred by the unperturbed triangular droplet crystal. The external potential therefore acts repulsively on the intrinsic droplets, pushing density away from its natural equilibrium sites. Moreover, the bipartite nature of the honeycomb lattice introduces additional intra-unit-cell frustration. Together, these effects generate substantially stronger geometric frustration than in the triangular case and make branch competition much more important.

\begin{figure}[tbp]
    \centering
    \includegraphics[width=0.99\linewidth]{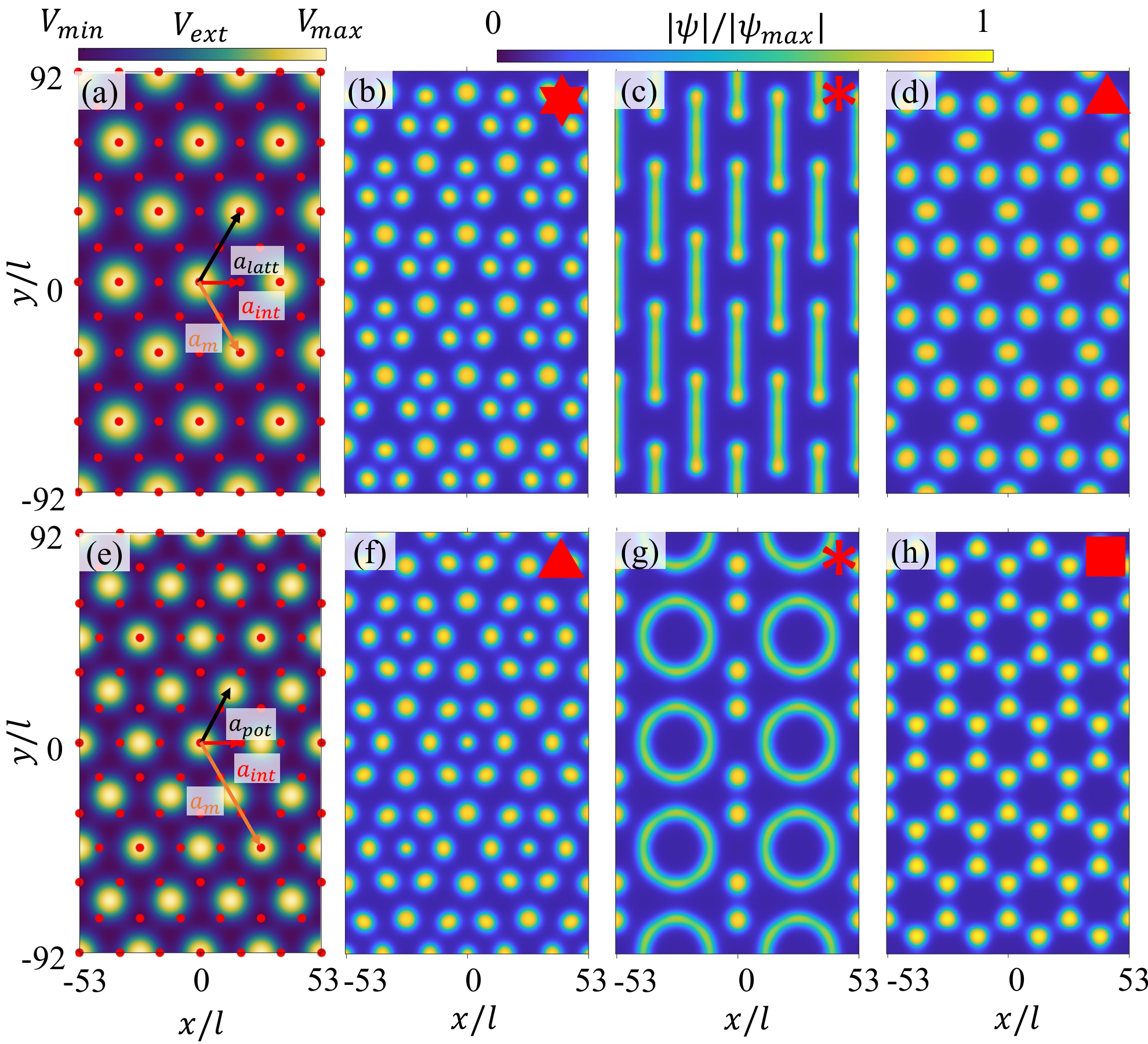}
    \caption{Ground-state density evolution of the dipolar Bose gas in a honeycomb optical lattice for $R < 1$.
    (a), (e) Superposition of the intrinsic triangular density (red dots) and the external potential (color map) for $R = 1/2$ and $R = 2/3$, respectively.
    The arrows indicate the three length scales $a_{\rm int}$ (red), $a_{\rm pot}$ (dark), and $a_m$ (orange).
    Unlike the triangular lattice, the potential maxima coincide with the intrinsic droplet positions.
    (b)--(d) Density profiles for $R = 1/2$ at $V_0 = 0.02$, $0.025$, and $0.027$, respectively.
    (f)--(h) Density profiles for $R = 2/3$ at $V_0 = 0.0075$, $0.01$, and $0.025$, respectively.
    The intermediate panels (c) and (g) display a symmetry-broken stripe phase and a ring-droplet lattice, respectively.
    In each density panel, the geometric marker identifies the initial state yielding the ground-state density, using the same symbol convention as in Fig.~\ref{fig2}.}
    \label{fig5} 
\end{figure}

For $R<1$, the competition between the repulsive honeycomb maxima and the intrinsic triangular order generates two distinct reconstruction pathways. For $R=1/2$ [Figs.~\ref{fig5}(a)--\ref{fig5}(d)], the weak-lattice state already differs qualitatively from the triangular-lattice case: each intrinsic droplet is suppressed at its center and split into two nearby maxima displaced toward lower-potential regions. Although the state remains globally $C_6$ symmetric, the density is now organized by repulsion from the honeycomb maxima rather than attraction to the minima. At intermediate depth, the system forms stripe segments along the honeycomb valleys, explicitly breaking rotational symmetry. At larger $V_0$, these stripes fragment into a discrete droplet array aligned with the honeycomb potential. The evolution thus follows a repulsion-driven pathway from droplet splitting to stripe formation and eventual lattice locking.

For $R=2/3$ [Figs.~\ref{fig5}(e)--\ref{fig5}(h)], the larger moir\'e period changes the preferred reconstruction. The weak-lattice regime still supports a smoothly modulated triangular superfluid, but with increasing $V_0$ the optimal branch changes and the intermediate state becomes a ring-droplet lattice rather than a stripe phase: the density reorganizes into annular droplets surrounding the low-potential honeycomb plaquettes. At larger $V_0$, the annuli break into localized peaks centered at the minima of the external field, yielding a honeycomb-controlled array. Relative to $R=1/2$, the ring-droplet state provides a smoother route into the deep-lattice regime.

\begin{figure}[tbp]
    \centering
    \includegraphics[width=0.99\linewidth]{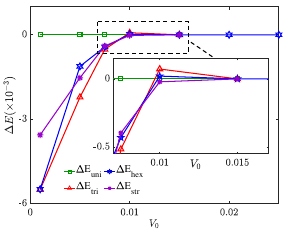} 
    \caption{{Energies of converged stationary branches obtained from different initial states for the system subjected to a honeycomb potential, with a fixed periodicity ratio $R=2/3$.} 
    Each curve corresponds to a different initial configuration including uniform (uni), triangular (tri), hexagonal (hex), and stripe (str).
    Energies are measured relative to the relaxed uniform state, so $\Delta_{\rm uni}=0$.}
    \label{fig6}
\end{figure}

The energetic origin of this behavior is illustrated in Fig.~\ref{fig6} for $R=2/3$. At weak lattice depth, the low-energy states remain connected to 2D crystalline initial conditions. With increasing $V_0$, however, the ground-state branch changes, reflecting a genuine reorganization of the density landscape as the honeycomb-induced strain overcomes the intrinsic triangular ordering tendency.

\begin{figure}[tbp]
    \centering
    \includegraphics[width=0.99\linewidth]{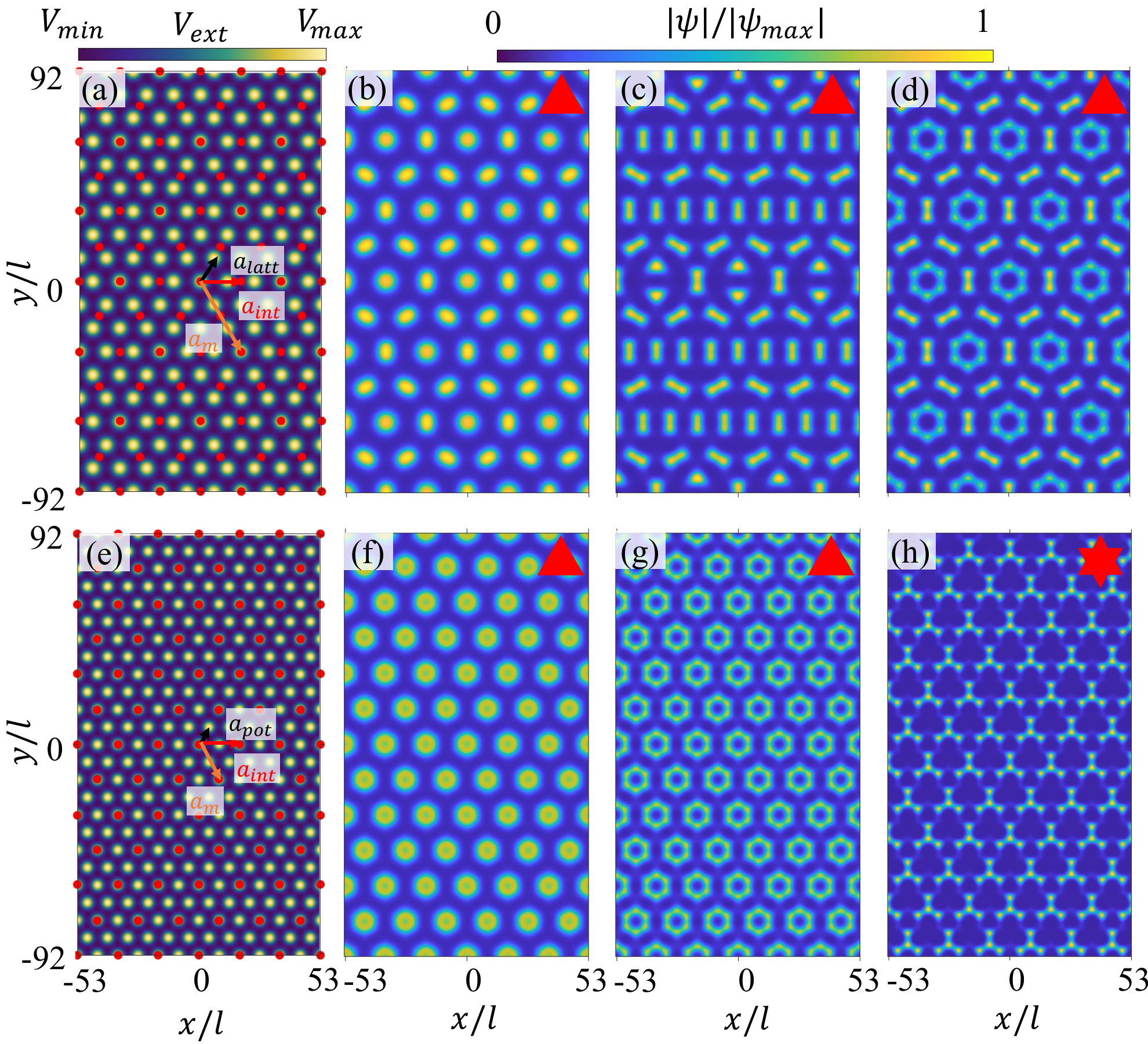}
    \caption{Ground-state density evolution of the dipolar Bose gas in a honeycomb optical lattice for $R > 1$.
    (a), (e) Superposition of the intrinsic triangular density (red dots) and the external potential (color map) for $R = 3/2$ and $R = 2$, respectively.
    The arrows indicate the three length scales $a_{\rm int}$ (red), $a_{\rm pot}$ (dark), and $a_m$ (orange).
    (b)--(d) Density profiles for $R = 3/2$ at $V_0 = 0.005$, $0.05$, and $0.1$, respectively.
    (f)--(h) Density profiles for $R = 2$ at $V_0 = 0.025$, $0.1$, and $0.5$, respectively.
    At higher potential depths, the density reorganizes into closed hexagonal loop structures in (d) and hexagonal ring droplets in (g) for $R = 3/2$ and $R = 2$, respectively.
    The deep-lattice limit in (h) reveals a honeycomb-commensurate superfluid in which the density forms a clover-like network aligned with $V_{\rm ext}$.
    In each density panel, the geometric marker identifies the initial state yielding the ground-state density, using the same convention as in Fig.~\ref{fig2}.}
    \label{fig7}
\end{figure}

For $R>1$, the honeycomb potential acts as a sub-droplet-scale repulsive corrugation. In this regime, the supersolid is more resistant to complete lattice locking than for $R<1$. For $R=3/2$ [Figs.~\ref{fig7}(b)--\ref{fig7}(d)], the triangular branch remains the lowest-energy branch throughout the range shown. The weak-lattice state is a moir\'e-modulated superfluid with two distinct local environments: droplets near honeycomb maxima are broadened and displaced, whereas interstitial droplets are elongated along the local valleys. At intermediate depth, the repelled droplets split into pairs and the interstitial density stretches into short segments, producing a symmetry-broken intermediate texture. At larger $V_0$, the density closes into hexagonal loops surrounding the honeycomb plaquettes.

For $R=2$ [Figs.~\ref{fig7}(f)--\ref{fig7}(h)], the same mechanism produces an even stronger reconstruction. The weak-lattice state remains moir\'e modulated, but the larger mismatch expands the moir\'e cell. At intermediate depth, the density around the repulsive maxima is almost completely expelled and reorganizes into a regular array of hexagonal ring droplets. At larger $V_0$, the system changes branch and develops a clover-shaped honeycomb superfluid network, in which each density maximum acquires three lobes extending along the honeycomb bonds. Reaching the deep-lattice regime requires substantially larger $V_0$ than in the $R<1$ cases, reflecting the substantial rigidity of the intrinsic density configuration against strong geometric mismatch.

\subsection{Heterostructural case II: square potential}

The square lattice produces the strongest symmetry mismatch considered here. A triangular droplet crystal cannot be made simultaneously commensurate with a square lattice along both spatial directions because the triangular geometry contains the irrational factor $\sqrt{3}$ relating the two axes. As a result, some degree of frustration persists across the entire parameter range. In the rigid-lattice limit, the superposition of triangular and square periodicities would be quasiperiodic. In the present soft system, however, the dipolar supersolid can deform continuously and thereby select compromise states. To verify that the resulting patterns are intrinsic rather than finite-size artifacts, we also performed enlarged-cell calculations for $R<1$, as discussed in Appendix~\ref{app:large-size}.

\begin{figure}[tbp]
    \centering
    \includegraphics[width=0.99\linewidth]{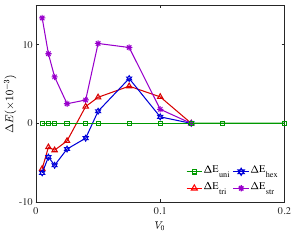}
    \caption{{Energies of converged stationary branches obtained from different initial states for the system subjected to a square potential, with a fixed periodicity ratio $R=1/2$.} 
    Each curve corresponds to a different initial configuration including uniform (uni), triangular (tri), hexagonal (hex), and stripe (str).
    Energies are measured relative to the relaxed uniform state, so $\Delta_{\rm uni}=0$.}
    \label{fig8}
\end{figure}

The representative energy landscape at $R=1/2$ is shown in Fig.~\ref{fig8}. At weak lattice depth, the triangular branch remains lowest in energy, consistent with a weakly perturbed supersolid. With increasing $V_0$, the ground state crosses over to a branch connected to the uniform initial state, indicating that the square confinement progressively suppresses the intrinsic triangular modulation and favors density patterns selected primarily by the external lattice.

\begin{figure}[tbp]
    \centering
    \includegraphics[width=0.99\linewidth]{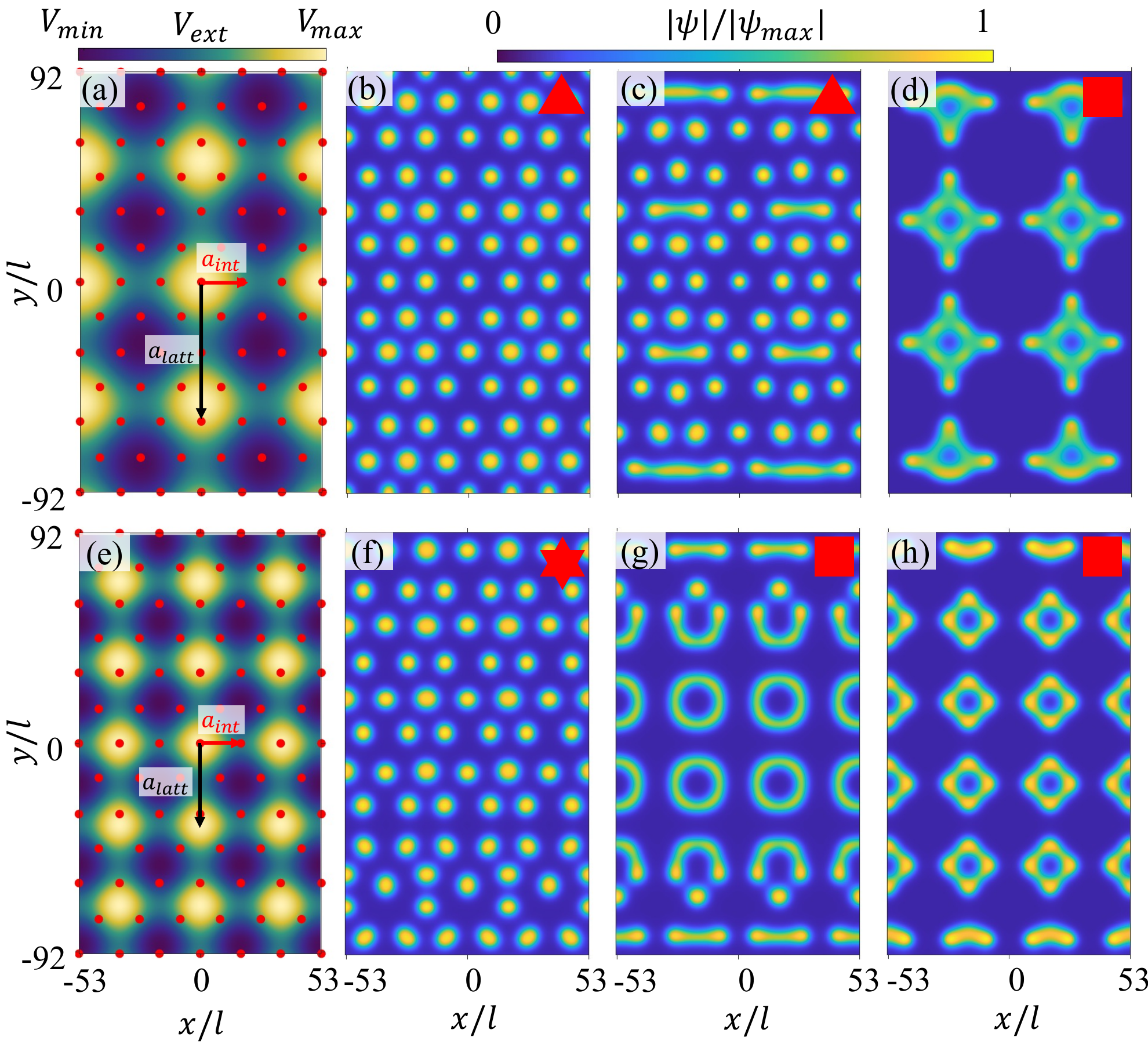}
    \caption{Ground-state density evolution of the dipolar Bose gas in a square optical lattice ($R < 1$).
    (a), (e) Superposition of the intrinsic triangular density (red dots) and the external potential (color map) for commensurability ratios $R = 1/3$ and $R = 1/2$, respectively.
    The arrows indicate the three length scales $a_{\rm int}$ (red) and $a_{\rm pot}$ (dark).
    (b)--(d) Density profiles for $R = 1/3$ at increasing lattice depths $V_0 = 0.05$, $0.075$, and $0.25$, respectively. Panel (c) shows a symmetry-broken stripe phase, whereas panel (d) shows a four-petaled fourfold-symmetric structure.
    (f)--(h) Density profiles for $R = 1/2$ at increasing lattice depths $V_0 = 0.025$, $0.05$, and $0.1$, respectively. The intermediate regime (g) exhibits a square array of ring-like droplets, while the deep-lattice limit (h) reveals compact diamond-shaped clusters.
    Panels (d) and (h) show ground states with $C_4$ symmetry, reflecting the symmetry of the external square potential. In each density panel, the geometric marker identifies the initial state yielding the ground-state density, using the same convention as in Fig.~\ref{fig2}.}
    \label{fig9}
\end{figure}

For $R<1$, the square lattice rapidly transfers its $C_4$ symmetry to the density distribution. For $R=1/3$ [Figs.~\ref{fig9}(b)--\ref{fig9}(d)], the weak-lattice state is still a moir\'e-modulated triangular superfluid with suppressed density near the square-lattice maxima. At intermediate depth, geometric frustration drives a stripe-like reconstruction oriented along the $x$ direction, leaving only mirror symmetry about the $y$ axis. At larger $V_0$, the {lowest-energy branch changes to the one connected to the uniform state} and the density is expelled from the repulsive maxima into four-petaled flower-like structures centered between neighboring maxima. The evolution thus realizes a crossover from intrinsic $C_6$ order to the $C_4$ symmetry of the square potential.

For $R=1/2$ [Figs.~\ref{fig9}(f)--\ref{fig9}(h)], the weak-lattice state is again moir\'e modulated. At intermediate depth, however, the uniform branch becomes energetically favorable and the density reorganizes into a regular square array of closed ring-like droplets encircling the square-lattice maxima. At larger $V_0$, these rings are compressed into compact diamond-shaped clusters oriented at $45^\circ$ with respect to the square axes. Compared with the triangular and honeycomb cases, the square lattice therefore produces a more direct symmetry transition from $C_6$ to $C_4$.

\begin{figure}[tbp]
    \centering
    \includegraphics[width=0.99\linewidth]{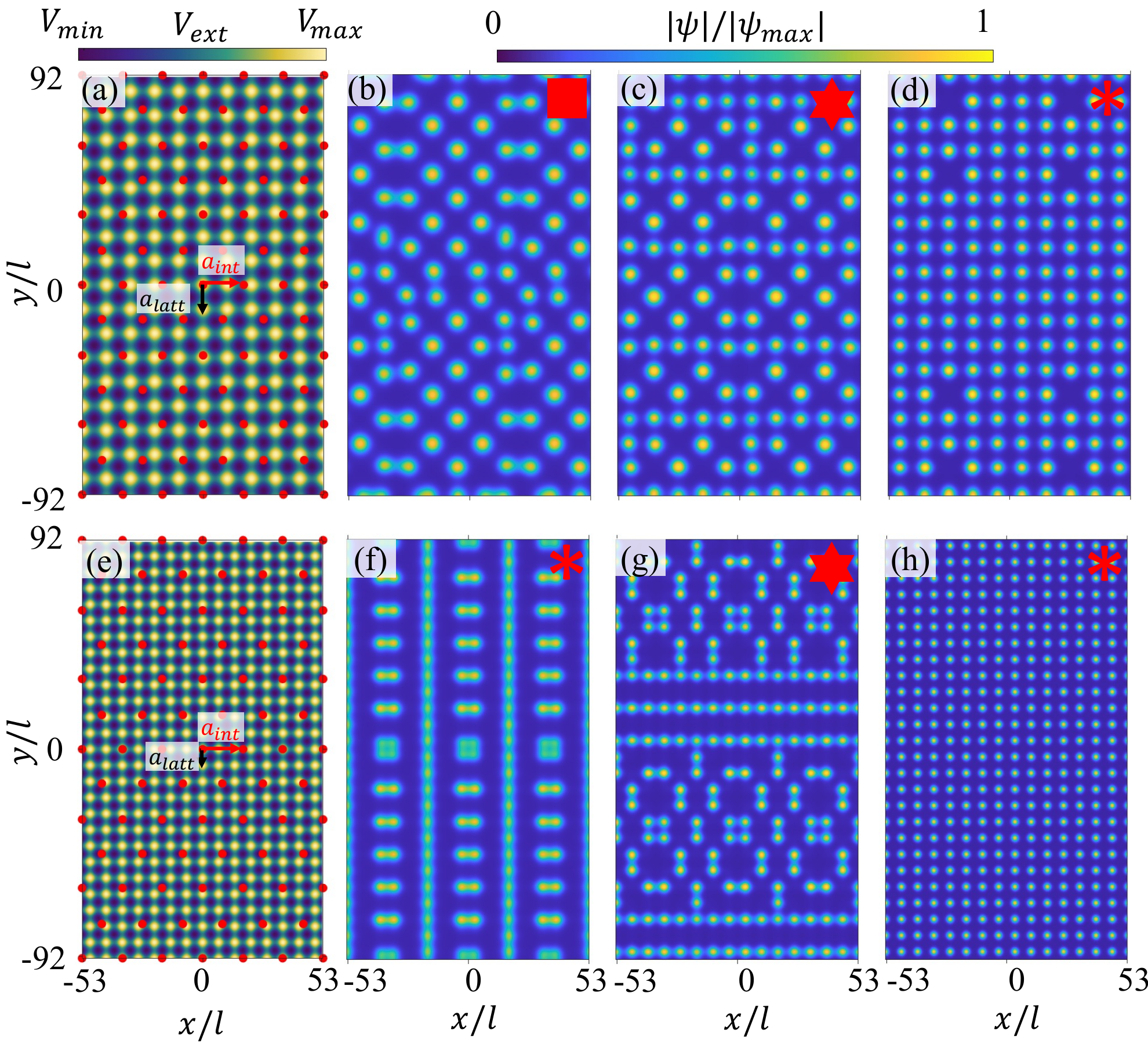}
    \caption{Ground-state density evolution of the dipolar Bose gas in a square optical lattice ($R > 1$).
    (a), (e) Superposition of the intrinsic density (color map) and the external potential (contours) for commensurability ratios $R = 5/3$ and $R = 5/2$, respectively.
    The arrows indicate the three length scales $a_{\rm int}$ (red) and $a_{\rm pot}$ (dark).
    (b)--(d) Density profiles for $R = 5/3$ at increasing lattice depths $V_0 = 0.01$, $0.025$, and $0.05$, respectively. Panel (c) shows a symmetry-broken stripe-like state before panel (d) approaches a discrete square-commensurate superlattice.
    (f)--(h) Density profiles for $R = 5/2$ at increasing lattice depths $V_0 = 0.025$, $0.1$, and $0.5$, respectively. The sequence evolves from a {clearly developed stripe-like state} in (f), through a modulated two-dimensional state in (g), to a cluster-crystal configuration in (h).
    In each density panel, the geometric marker identifies the initial state yielding the ground-state density.
    In each density panel, the geometric marker identifies the initial state yielding the ground-state density, using the same symbol convention as in Fig.~\ref{fig2}.}
    \label{fig10}
\end{figure}

For $R>1$, the intrinsic droplet spacing exceeds the square-lattice period, and the external field becomes a dense sub-droplet corrugation. Figure~\ref{fig10} shows that this regime supports robust anisotropic intermediate states. For $R=5/3$ [Figs.~\ref{fig10}(b)-\ref{fig10}(d)], the weak-lattice state already exhibits elongated droplets because the dense square modulation generates valley-like channels between adjacent maxima. At intermediate depth, these elongated structures sharpen into a more pronounced stripe texture. At larger $V_0$, the density fragments into a discrete square-commensurate superlattice with residual vacancies, showing that {the density becomes primarily shaped by the external lattice}.

The case $R=5/2$ [Figs.~\ref{fig10}(f)--\ref{fig10}(h)] follows the same overall trend but requires substantially larger $V_0$ because of the stronger mismatch. The weak-lattice state is a clearly developed stripe-like state aligned with the square axes. At intermediate depth, the lowest-energy branch changes to a more complex 2D configuration that retains partial stripe character while developing transverse modulations. At larger $V_0$, the stripe branch again becomes favorable and the system enters a cluster-crystal regime in which multiple density maxima occupy each square potential well. Thus, in the square lattice, frustration is relieved only gradually, through a sequence of strongly anisotropic states, before the system finally reaches a lattice-dominated configuration.

\section{Conclusion}
In this work, we have investigated the ground-state structures of a 2D dipolar supersolid subjected to triangular, honeycomb, and square optical lattices by means of imaginary-time evolution of eGPE with multiple initial-state configurations. By varying the lattice depth $V_0$ and the commensurability ratio between the intrinsic droplet spacing and the external lattice period, we identified multiple density reconstructions arising from the competition between interaction-driven crystallization and rigid optical confinement.

The structural response is governed primarily by symmetry compatibility and commensurability. Across all lattice geometries, a common feature is the emergence of symmetry-broken intermediate states. These include stripe-like and cluster-like configurations that interpolate between the weak-lattice moir\'e-modulated superfluid and the strongly pinned deep-lattice phases. Their appearance indicates a complex energy landscape with multiple competing metastable minima. More generally, increasing incommensurability systematically shifts the onset of the deep-lattice regime to larger $V_0$, demonstrating that the interaction-stabilized supersolid resists external locking more strongly when the two characteristic length scales are poorly matched.

The heterostructural cases further highlight two distinct frustration mechanisms. For the honeycomb lattice, the coincidence of intrinsic droplet positions with repulsive regions of the optical potential promotes droplet splitting, loop-like structures, and other reconstructed states. For the square lattice, the irreducible mismatch between the intrinsic $C_6$ order and the imposed $C_4$ symmetry leads to rotational-symmetry breaking and ultimately to lattice-commensurate cluster-crystal or discrete-superlattice states. These results show that optical lattices provide a means of controlling not only the periodicity but also the symmetry and connectivity of dipolar supersolid order.

Taken together, these findings broaden the scope of moir\'e physics by revealing an unconventional setting in which moir\'e superstructures emerge from the interplay between spontaneous crystallization and external periodic confinement. This interplay between a soft self-organized crystal and a rigid external lattice provides a setting for studying frustrated quantum states in dipolar gases. It would be interesting in future work to examine the superfluid response, collective excitations, and defect dynamics of these modulated phases, as well as the effects of temperature and quantum depletion beyond the present mean-field framework. Given current experimental capabilities in dipolar Bose gases such as $^{164}$Dy and $^{166}$Er, the phases predicted here should be accessible with existing optical-lattice techniques \cite{lu2011strongly,aikawa2012erbium,baier2016extended,su2023dipolar,norcia2021two,sohmen2021birth}.
\appendix

\section{Large-size simulations}
\label{app:large-size}
\begin{figure}[tbp]
    \centering
    \includegraphics[width=0.99\linewidth]{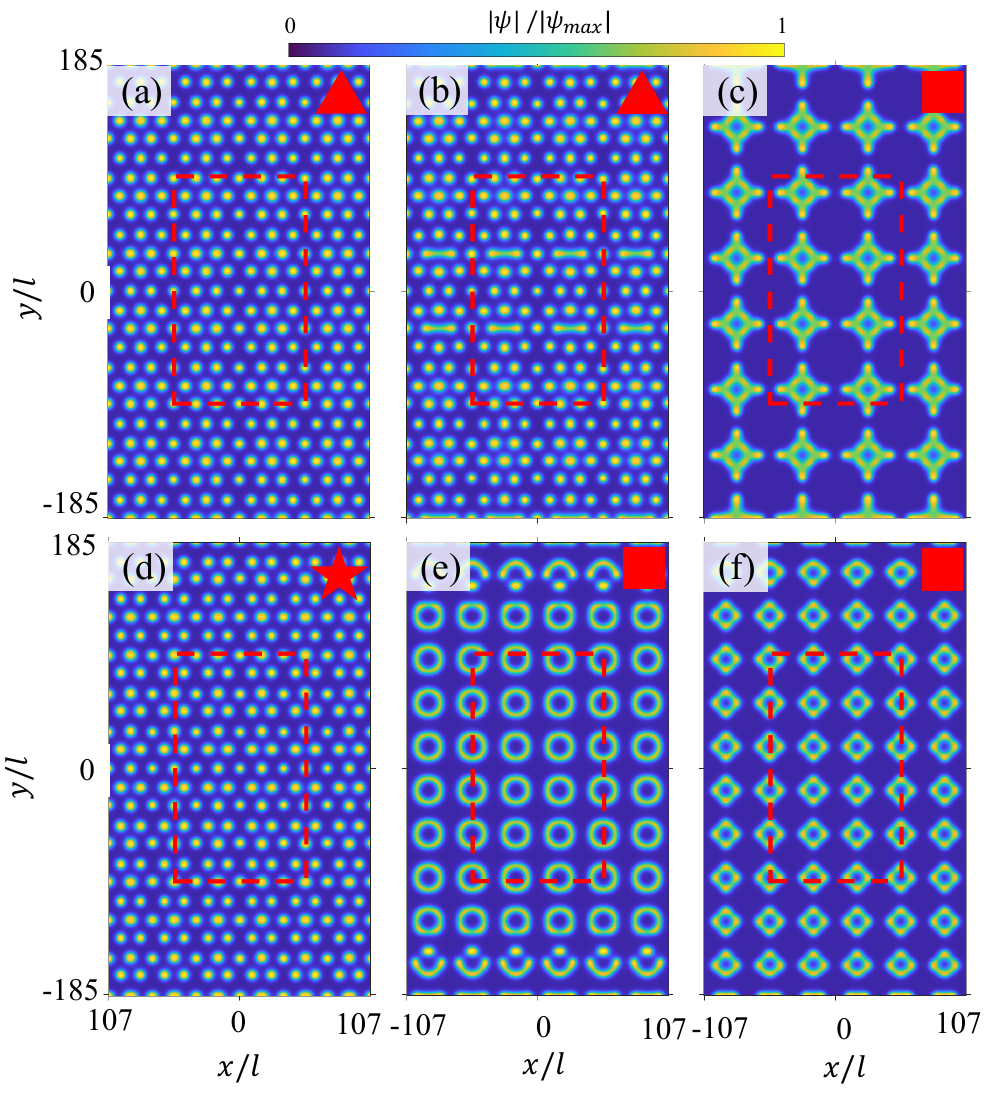}
    \caption{Ground-state density profiles of the dipolar Bose gas in a square optical lattice, computed in a simulation cell approximately twice as large as that used in Fig.~\ref{fig9} to assess finite-size effects for $R<1$. The panels correspond to the same parameter sets as in Fig.~\ref{fig9}: (a)--(c) $R=1/3$ at $V_0=0.05$, $0.075$, and $0.25$, respectively; (d)--(f) $R=1/2$ at $V_0=0.025$, $0.05$, and $0.1$, respectively. Red dashed rectangles mark the extent of the smaller simulation cell used in the main text. The persistence of the same density motifs inside and outside these rectangles demonstrates that the reported structures are bulk phases rather than boundary-induced artifacts.
    In each density panel, the geometric marker identifies the initial state yielding the ground-state density, using the same symbol convention as in Fig.~\ref{fig2} of the main text.}
    \label{fig11}
\end{figure}

In the main text, the square-lattice results for $R<1$ were obtained in a finite computational cell [Fig.~\ref{fig9}]. Because the square lattice is incommensurate with the intrinsic triangular droplet crystal, one must verify that the symmetry-broken intermediate states reported there are genuine bulk configurations rather than artifacts of the chosen cell size or boundary matching. To address this point, we repeated the calculations in a simulation domain that is approximately twice as large as that used in Fig.~\ref{fig9}. Specifically, we have increased $L_x = 6a_{\mathrm{int}}$ and $L_y = 6\sqrt{3}\,a_{\mathrm{int}}$, and discretized the enlarged computational domain on a $512 \times 512$ grid. Since both the system size and the number of grid points were doubled along each direction, the spatial grid spacing remained unchanged. This allows us to assess finite-size effects without altering the spatial resolution.

The resulting density profiles are shown in Fig.~\ref{fig11} for the representative cases $R=1/3$ and $R=1/2$. The red dashed rectangles indicate the extent of the smaller simulation cell employed in the main text. In all cases, the density patterns inside these rectangles coincide with those obtained in Fig.~\ref{fig9}, while the same structures continue periodically throughout the enlarged domain. This agreement shows that the weak-lattice superfluid states, the intermediate symmetry-broken configurations, and the deep-lattice fourfold-symmetric localized phases are not induced by the system boundaries.

We therefore conclude that the structural sequences reported in the main text for the square optical lattice, including the symmetry-broken intermediate regimes, are robust against enlargement of the simulation cell and faithfully represent the bulk behavior of the dipolar supersolid.

\begin{acknowledgments}
This work was supported by the National Key Research and Development Program of China (Grant No. 2022YFA1405304), the National Natural Science Foundation of China (Grant No. 12574193) and the Guangdong Provincial Quantum Science Strategic Initiative (Grant Nos. GDZX2401002 and GDZX2501003).
\end{acknowledgments}
\bibliography{refs}
\end{document}